\documentclass[prl,twocolumn,floatfix,superscriptaddress]{revtex4}
\usepackage{dcolumn,amsmath}
\usepackage[dvips]{graphicx}
\usepackage{bm}
\usepackage{hyperref}  

\usepackage{hhline}
\usepackage{slashbox}

\newcommand{\Eeff}{\ensuremath{E_{\rm eff}}}
\newcommand{\Eext}{\ensuremath{E_{\rm ext}}}
\newcommand{\eEDM}{{\em e}EDM}
\newcommand{\ecm}{\ensuremath{e {\cdotp} {\rm cm}}}

\begin{document}
 \title{Towards the search of electron electric dipole moment: correlation calculations of the
        P,T-violation effect in the Eu$^{++}$ cation.}
\author{L.V.\ Skripnikov}\email{leonidos239@gmail.com}
\author{A.V.\ Titov} 
\author{A.N.\ Petrov}
\altaffiliation [Also at ] {St.-Petersburg State University, St.-Petersburg,
        Russia}
\author{N.S.\ Mosyagin}
\affiliation
{Petersburg Nuclear Physics Institute, Gatchina,
             Leningrad district 188300, Russia}
\author{O.P.\ Sushkov}
\affiliation{School of Physics, University of New South Wales,
             Sydney 2052, Australia}
\date{26 March 2011}

\begin{abstract}
Recently the Eu$_{0.5}$Ba$_{0.5}$TiO$_{3}$ solid was suggested as a promising
candidate for experimental search of the electron electric dipole moment.  To
interpret the results of this experiment one should calculate the
effective electric field acting on an unpaired (spin-polarized) electrons of
europium cation in the crystal because the value of this field cannot be
measured experimentally.  The Eu$^{++}$ cation is considered in the paper in the
uniform external electric field \Eext\ as our first and simplest model
simulating the state of europium in the crystal.  We have performed high-level
electronic structure correlation calculation using coupled clusters theory
(and scalar-relativistic approximation for valence and outer core electrons at
the molecular pseudopotential calculation stage that is followed by the
four-component spinor restoration of the core electronic structure)
to evaluate the enhancement coefficient $K= \Eeff/\Eext$ (where  \Eext\ is the
applied external electric field and \Eeff\ is the induced effective electric
field acting on an unpaired electron in Eu$^{++}$).  A detailed computation
analysis is presented. The calculated value of $K$ is -4.6.
\end{abstract}

\maketitle

\section{Introduction}

During the past decades a significant experimental and theoretical efforts have
been undertaken to measure an electric dipole moment of the electron (\eEDM\ or
$d_e$ below). \eEDM\ is of fundamental importance for theory of P,T-odd
interactions because its existence violate both space parity (P) and time
reversal (T) symmetries \cite{Khriplovich:97,Commins:98}. The Standard model
prediction for \eEDM\ is of order $10^{-38}\ecm$ of magnitude or even less, but
the most of other modern theoretical models predict much higher values,  on the
level of $10^{-27} - 10^{-29}\ecm$ \cite{Commins:98}. Current experimental
upper bound for {\em e}EDM is obtained in the measurements on the atomic Tl beam
\cite{Regan:02} and constitutes $1.6  \cdotp 10^{-27}\ecm$. 
Therefore, increasing the experimental sensitivity on even one-two orders of magnitude for
the value of {\em e}EDM will dramatically influence all the popular models
suggesting a ``new physics'' beyond the Standard model, in particular
supersymmetry, even if bounds on the P,T-odd effects compatible with zero are
obtained (see \cite{Ginges:04, Erler:05} and references therein). 

Nowadays there are several experimental setups that are using molecules
containing heavy atoms to measure {\em e}EDM. These include neutral molecule
experiments, e.g., the beam experiment on the YbF molecular radicals carried on
by Hinds and co-workers \cite{Sauer:06a}; another one employs vapor cell in
experiment on the metastable $a(1)$ state of PbO that is prepared by the group
of DeMille (see \cite{Kawall:04b, Bickman:09} and references therein), the
Stark-trap experiment on the PbF radicals is prepared by Shafer-Ray
\cite{Shafer-Ray:08, Shafer-Ray:09} and some new beam experiments are now
prepared on the metastable $^3\Delta_1$ state of ThO* \cite{Vutha:2010} and the ground
$^3\Delta_1$ state of WC.  In the other series of experiments suggested by
Cornell and co-workers some trapped cold molecular cations are planned to be
used. Up to now several cations were considered including HI$^+$, HfF$^+$,
PtH$^+$, ThF$^+$, etc.\ (see \cite{Leanhardt:10E} and references).

The ideas to use solids for such experiments were proposed by Shapiro many yeas
ago \cite{Shapiro:1968}. However, only during the last decade such experiments
to search for \eEDM\ have become attractive due to suggestions of Lamoreaux
\cite{Lamoreaux:02} and Hunter \cite{Hunter:01} to use GdGaO and GdFeO. Recently
a new kind of solid-state experiment was proposed on the
Eu$_{0.5}$Ba$_{0.5}$TiO$_{3}$ (EBTO) crystal \cite{Suskov:10} having perovskite
structure. In this crystal Eu has seven unpaired spin-aligned electrons in
4f-shell and, therefore, nonzero magnetic moment. Besides, EBTO has
ferroelectric phases at low temperatures
\cite{Rushchanskii:10}. This experiment (as well as the atomic and molecular
experiments) employs the idea \cite{Landau:57} that the electron EDM has to
point along its magnetic moment (spin). As a result, when an electric field,
\Eext, is applied to a sample lifting the degeneracy between electrons with EDMs
parallel and antiparallel to \Eext, the associated imbalance of electron
populations generates a magnetization \cite{Rushchanskii:10}. The orientation of
the magnetization is reversed when the electric field direction is switched; and
this change in sample magnetization is supposed to be monitored using a SQUID
magnetometer. 

However, to extract the value of $d_e$ it is necessary to know the value of the
effective electric field acting on the unpaired electrons of Eu in
Eu$_{0.5}$Ba$_{0.5}$TiO$_{3}$ \cite{Suskov:10} which can not be measured. This
is very difficult computational problem even for a solitary Eu$^{++}$ cation
(see below) and this paper starts our {\it ab~initio} studies of \Eeff\ in the
Eu$_{0.5}$Ba$_{0.5}$TiO$_{3}$ crystal.  Here we consider the computationally
simplest model simulating the electronic structure of Eu in EBTO: Eu$^{++}$
cation in an external electric field.

\section{Methods}

When an atom (ion) with the unpaired electrons is placed into the external
electric field \Eext, the resulting effective field \Eeff\ acting
on an unpaired electron is proportional to the applied (weak-) field with the
enhancement coefficient $K$:

$$\Eeff=K \cdotp \Eext$$
It was Sandars who discovered that P,T-odd effects can be
strongly enhanced in heavy atoms due to the relativistic effects
\cite{Sandars:65}.
A very useful semiempirical expression for $K \sim \alpha^2 Z^3$ is proposed in \cite{Flambaum:76} 
that is well working for $s, p$ electrons though it is questionable for $f$ electrons.
\Eeff\ is given by \cite{Kozlov:87, Kozlov:95, Titov:06amin}):
\begin{equation}
 \label{matrelem}
   E_{\rm eff}= \langle \Psi|\sum_iH_d(i)|\Psi \rangle,
\end{equation}
\begin{eqnarray}
 \label{PTOperator}
    H_d(i)=2d_e
    \left(\begin{array}{cc} 0 & 0 \\
     0 & \bm{\sigma_{\rm i} E({\bf r}_{\rm i}}) \\
    \end{array}\right)\ ,
\end{eqnarray}
$\Psi$ is the wave function of the atom (ion) in an external electric field
\Eext; $\bm{E}({\bf r}_i)$ is the sum of electric fields from nuclei
and electrons.  The  wavefunction must  take account of
the most part of the relativistic and relevant correlation effects (see below) of valence
  (and sometimes outer-core) electrons. These electrons are the most affected
%
(polarized) by the applied electric field and dramatically influence on \Eeff.
The polarization of the inner-core electrons is usually negligible. These
circumstances allow us to use a two-step technique, advanced by our group
\cite{Titov:96, Mosyagin:98, Petrov:02, Petrov:05a, Titov:06amin} and recently
applied for calculation of \Eeff\ in molecular systems \cite{Skripnikov:09,
Petrov:07a}. {\em At the first step} we exclude inactive inner-core orbitals
from correlation calculation with the help of a very accurate generalized
relativistic effective core potential method (GRECP)
\cite{Mosyagin:00,Isaev:00,Mosyagin:01b} to reduce computational efforts.
Besides, the valence orbitals are smoothed in cores and this smoothing allows
one to reduce the number of primitive Gaussian basis functions required for
appropriate description of valence spinors in subsequent molecular calculations.
Moreover, as an alternative to a four-component calculation with small
components of Dirac bispinors one can perform two- or one-component calculation,
i.e., with or without spin-dependent interactions (which can include Breit etc.\
terms additionally to the spin-orbit ones) for explicitly treated electron shells
taken into account.  This procedure dramatically reduces computation time with
minimal loss of accuracy (the particular choice of the valence~/~core electron's
partitioning is dependent first of all on the electronic structure of considered
system and then, on the accuracy required).  It should be noted that the
all-electron four-component calculations are much more time and resource
consuming for the same level of accuracy as highly accurate GRECP ones.

{\em At the second step}, when the GRECP calculation is performed (with or
without accounting for the electron correlation), we restore the four-component
core electronic structure of valence orbitals and, thus, the corresponding
relativistic one-electron density matrix from the GRECP one.  Using the
restored density matrix one can easily calculate any one-electron properties
(such as \Eeff, hyperfine constants, etc.) which have the most contributions in
the atomic core regions.  We have chosen the coupled clusters (CC) approach as
the main instrument to account for electron correlation because it has a number
of advantages over the other methods (such as restricted active space SCF (RASSCF),
configuration interaction (CI), and many-body perturbation theory, which were also
used at the preliminary stage) for our purposes: rather quick convergence of the
results for the property of our interest with increasing the level of accounting
for the electron excitations; well suppressed spin contamination problem (this
is especially valuable for Eu$^{++}$ having seven unpaired electrons
with the multiplicity equal to 8).

Finally, the most important aspect of the present calculation should be
stressed.  The operator (\ref{PTOperator}) is nonzero only between the states of
opposite parities, e.g., $s-p, p-d, f-d$, etc. Moreover, the corresponding
molecular orbitals (on which this mixing takes place) should be spin-polarized
or singly-occupied, otherwise, the contribution from a ``spin-up'' matrix
element is completely compensated by some corresponding ``spin-down'' matrix
element.  The Eu$^{++}$ cation has the ground state electronic configuration
$[..]4s^2 4p^6 4d^{10} 5s^2 5p^6 4f^7_{\uparrow}$.  Therefore, in an external
electric field one should expect polarization of $4f_{\uparrow}$ unpaired
electrons into unoccupied low-lying $5d$ states.  However such a mixing will not
result in a big value of the enhancement coefficient $K$ because $4f$ and $5d$
have very small amplitudes in the vicinity of the Eu nucleus, where the operator
(\ref{PTOperator}) (mainly the electric field from the Eu nucleus) is big.  The
$K$ value magnitude a few orders higher may be expected in the case of $s-p$
mixing which takes place, e.g., in heavy alkali metals, such as cesium with
$[Xe]6s^1$ configuration, or in $p^1$-elements like thallium with $[Hg]6p^1$
configuration (or in a number of molecules such as HfF$^+$ in the $^3\Delta_1$
state where the $s-p$ mixing is very large due to the internal structure
asymmetry of polar molecules. However, Eu$^{++}$ has no unpaired electrons in
$s$ or $p$ states and contribution to $K$ from the matrix element of operator \ref{PTOperator} due
to the spin-polarized mixing of these states may occur only at the third and
higher orders of perturbation theory, when both the weak (P,T-odd) and external
field are treated as perturbations together with the electron correlation
effects \cite{Buhmann:02}.
Thus, though the $s-p$ matrix elements of operator (\ref{PTOperator}) are big,
the coefficients in front of these elements in the case of Eu$^{++}$ are to be
small and dramatically dependent on the quality of accounting for the electron
correlation (see \cite{Isaev:05a} as an example of strong influence of
correlation effects).  That is why the calculation of $K$ for Eu$^{++}$ is much
more complicated as compared to study of the mentioned atoms and molecules with
respect to the required quality of accounting for electron correlations.  We do
not account for the spin-orbit effects in this research and concentrate our
attention on the dynamic correlations since our estimates show that spin-orbit
contributions do not influence strongly on the \Eeff\ value whereas simultanious
treatment of both spin-orbit and correlation effects crucially reduce the
possibilities of accounting for important dynamic correlations (and can not be
currently taken into account at the correlation level exploited in the paper).

\section{Computational details}

For Eu$^{++}$ cation, 28 core electron GRECP (with $1s-3d$ electrons in core) was
generated and used for subsequent correlation calculations. In order to check a
reliability of our value for $K$ we have performed a detailed analysis of
approximations used in our calculations such as basis set completeness, required
level of accounting for correlation, etc., that is given below. To perform the
correlation calculations we have used MRCC \cite{Kallay:1,Kallay:3} and CFOUR
\cite{CFOUR} codes. To perform DFT calculations we have used 
US-GAMESS program package \cite{USGAMESS1}.

\subsection{Basis set generation}

For the Eu$^{++}$ cation the contracted correlation scheme of the basis set
generation from papers \cite{Mosyagin:00,Isaev:00,Mosyagin:01b} was used. This
scheme assumes that any last added function does not change the energy of the
most important transitions (describing the state of atom-in-a-molecule) more
than some threshold (10 cm$^{-1}$ in this paper).  The generated basis set
includes six s-type, seven p-type, five d-type, four f-type and two g-type
generally contracted Gaussian functions. To check the merits of the generated
basis set for evaluating \Eeff, a series of coupled clusters (CC) calculations
with single and double amplitudes (CCSD) has been performed with increasing
step by step the number of basis functions. The results are given in table
\ref{Tbasis1a}.

\begin{table}[!h]
\caption{
$K$ values calculated using the CCSD method and different basis sets. ns, np,
nd, nf and ng - are the numbers of s-, p-, d-, f- and g- contracted Gaussian
functions included in the corresponding basis set
}
\label{Tbasis1a}
\begin{tabular}{ c  c  c  c  c  l }
\hline
\hline
  \textbf{ns} &
  \textbf{np} &
  \textbf{nd} &
  \textbf{nf} &
  \textbf{ng} &
  \bfseries K(CCSD) 
\\\hline
  2 &
  2 &
  4 &
  3 &
  0 &
  {}-0.9 
\\
  3 &
  3 &
  4 &
  3 &
  0 &
  1.3 
\\
  4 &
  5 &
  4 &
  3 &
  0 &
  2.3 
\\
  4 &
  5 &
  5 &
  4 &
  0 &
  2.2 
\\
  4 &
  5 &
  5 &
  4 &
  2 &
  2.2 
\\
  5 &
  6 &
  5 &
  4 &
  2 &
  0.1 
\\
  6 &
  6 &
  5 &
  4 &
  0 &
  {}-1.3 
\\
  6 &
  6 &
  4 &
  3 &
  0 &
  {}-1.2 
\\
  6 &
  7 &
  4 &
  3 &
  0 &
  {}-3.0 (*) 
\\
  6 &
  7 &
  5 &
  4 &
  0 &
  {}-3.1 (**)
\\
  6 &
  7 &
  5 &
  4 &
  2 &
  {}-2.8 
\\\hline
14 &
14 &
5 &
4 &
0 &
{}-4.1
%
(Lbas)
\\
20 &
20 &
5 &
4 &
0 &
{}-4.1\\
14 &
14 &
12 &
4 &
0 &
{}-4.4\\\hline\hline
\end{tabular}
\end{table}

It is clear from table \ref{Tbasis1a} that $K$ in the Eu$^{++}$ ion strongly
depends on the number of basis functions, e.g., even the latest added sixth and
seventh contracted correlation functions give significant contributions to $K$.
It is clear from the comparison of (*) and (**) lines that for d-type and f-type
functions one can keep only 4 and 3 functions, respectively. The inclusion of
$g-$type functions in calculation gives negligible contribution to $K$ (that is
well understandable in context of the discussed above $f-d$ mixing), therefore,
they can be completely excluded from the basis set. 

Accounting for the highest sensitivity of $K$ to $s$ and $p$ functions we have
performed calculations with uncontracted $s$ and $p$ functions (14 s- and 14 p-
primitive gaussians).  To check that the basis set is complete enough regarding
to evaluation of $K$, the calculations with 20 s- and 20 p- functions have also
been performed.  However, the magnitude of $K$ did not change.  The additional
uncontracting the $d-$orbitals (12 primitive $d-$type gaussian functions)
changes $K$ value only by ~7\%. As the computational time and resources for
highly correlation study, such as CC with single, double and triple amplitudes
CCSDT, strongly depends on the number of basis functions, we have chosen basis
set which contains 14 primitive s-gaussians, 14 primitive p-gaussians, 5
contracted d-gaussians and 4 contracted f-gaussians which give the ``converged''
value of K. Below we shall refer to this basis set as to LBas.

\subsection{Optimal external field strength}

To compute the enhancement factor $K$, the linear dependence of \Eeff\ with
respect to the applied field \Eext\ is required and the following two
circumstances have to be satisfied:
(i) the field must be strong enough to reduce influence of computational
errors (round-up etc.) on $K$;
(ii) the field must be weak enough to prevent significant perturbations of
the electronic structure (the physical external field is very small in practice
and only the first-order perturbation of the wave function, linear on \Eext,
should be taken into account).

We have performed a series of the CCSD calculations and have found that the
linearity is provided in a wide range including $10^{-6}-10^{-1}\ a.u.$ and
have chosen $\Eext{=}0.001~a.u.$ for our further calculations.

\subsection{Choosing the correlation method}

It is well known that the methods based on the unrestricted Hartree-Fock (UHF)
reference are not free from the spin contamination problem.  Therefore, we have
performed coupled clusters calculations also with the restricted open-shell
Hartree-Fock (ROHF) reference.  In this case the spin contamination problem
excluded at the level of the reference (but can arise at the coupled cluster
treatment stage due to features of the used codes).  Table \ref{TCorrMethod}
gives values of $K$ calculated at different levels of correlation treatment.

\begin {table}[!h]
\caption{
 Calculated $K$ values with different correlation methods using UHF and ROHF
 references. Mean values of the square of spin $\langle S^2 \rangle$ operator
 are given in brackets (for the octet multiplicity ``clean'' $\langle S^2
 \rangle=15.75$ ).
}
\label{TCorrMethod}
\begin{tabular}{ c  c  c }
\hline
\hline
~
 &
\multicolumn{2}{ c }{$K$}\\ \hhline{~--}
\backslashbox{method}{reference} &
\sffamily UHF [ $\langle S^2 \rangle$ ] &
\sffamily ROHF [ $\langle S^2 \rangle$ ]\\ \hline

\sffamily CCSD &
{\sffamily {}-4.1}

\textsf{[15.75033]} &
{\sffamily {}-4.6}

\textsf{[15.75026]}\\ 
\sffamily CCSDT &
{\sffamily {}-4.6}

\textsf{[15.75000]} &
{\sffamily {}-4.6}

\textsf{[15.75000]}\\ \hline
\sffamily MP2 &
 \sffamily {}-4.4 &
 \sffamily {}-3.6  
\\ 
\sffamily MP3 &
\sffamily {}-2.5 &
\sffamily {}-2.7 
~
\\ 
\sffamily MP4 &
\sffamily {}-5.5 &
 ~ --

\\\hline\hline
\end{tabular}
\end{table}

One can see from this table that accounting for the iterative triple amplitudes
within the UHF-CCSDT method increases the magnitude of $K$ by less than 15\% as
compared to $K$, calculated at the UHF-CCSD level.  ROHF-based CC methods give
$K=-4.6$ already at the CCSD level and inclusion of triples do not change $K$
value\footnote
  {To check the reliability of ROHF-CCSD we have also performed
  ``unrelaxed'' ROHF-CCSD calculation, i.e., with the ROHF-reference taken from
  zero \Eext\ field calculation. This have resulted in 23\% divergence of $K$
  from the $-4.6$ value.  However, the unrelaxed ROHF-CCSDT calculation (see the
  next section) give the same $K$ value as the relaxed UHF-CCSDT and
  ROHF-CCSDT.}

From the values of the mean squared spin operator one can see that the
spin-contamination problem is not dramatic already at the CCSD level and is
negligible at the CCSDT level.
All these facts are good arguments that our ultimate value for $K$, -4.6, is reliable.

Table \ref{TCorrMethod} also illustrates why we have chosen coupled clusters
method. The M\o ller-Plesset (MP) perturbation theory is not converged for $K$
even at the fourth order.  {\em From the other hand this illustrates that the
enhancement factor $K$ is determined by rather high-orders of perturbation
theory where Coulomb interaction between electrons (accounting for correlation)
is considered as perturbation}.

Summarizing, it follows from the above analysis that the use of the CCSDT method
is sufficient for reliable calculation of $K$ and our final value is $K=-4.6$.

 \section{Analysis  of contributions to $K$}

It was mentioned above that one of stages of calculating $K$ is evaluation of
the spin density matrix. Therefore it is possible to estimate contributions from
the mixing of the basis functions having different angular momenta by setting
all of the other elements of the spin density matrix to zero.  By this way we
have estimated such contributions to $K$ from the spin density matrix calculated
at the CCSDT level. The contributions are given in table
\ref{TPairContributions}:

\begin{table}[!h]
\caption{
Pair contributions to $K$ calculated at the CCSDT level.
}
\label{TPairContributions}
\begin{tabular}{ c | c  c  c  c  c  c }
\hline\hline
~
 &
s &
p &
d &
f\\\hline
s &
\centering {}0 &
\centering {}-3.3 &
\centering 0 &
0\\
p &
~
 &
\centering {}0 &
\centering +0.3 &
0\\
d &
~
 &
~
 &
\centering {}0 &
{}-1.6\\
f &
~
 &
~
 &
~
 &
{}0\\\hline\hline
\end{tabular}
\end{table}

It follows from this table that the main
   (cumulative)
contribution to $K$ is provided by the
$s-p$ mixing. The next important contribution (which is twice smaller) is
provided by the $f-d$ mixing. Contribution from the $p-d$ mixing is almost
negligible. 

It is very instructive to estimate contributions to $K$ from the individual
(outer-core and valence) $ns,np$ shells.  Qualitatively, the spin exchange of a
given $ns_{\uparrow}$ or $np_{\uparrow}$ orbital with $(4f_{\uparrow})^7$, lead
to spin-polarization of the shells and, being space-polarized by \Eext, to
uncompensated contribution to $K$. It is first dependent on the following two
factors: (i) space localization relative to the $4f-$shell (which is occupied by
seven alpha-spin electrons); (ii) energy separation relative to the polarizing
orbitals whose energies, by order of magnitude, are close to zero.  From table
\ref{TOrbitalParameters} it can be seen that $4s$ and $4p$ orbitals are
localized at the same region as $4f-$orbitals, at the same time last maxima of
the $5s$ and $5p$ orbitals are in about $1.5$ times larger. From the other hand,
$5s,5p$ energy factors (denominators) are much smaller than those for the
$4s,4p$. In turn, the space polarization $s-p$, $p-d$ etc.\ due to the \Eext\
should be expected notably stronger for the $5s$ and $5p$ orbitals whereas
matrix elements of (\ref{PTOperator}) are smaller for them.  Thus both $5s,5p$
as well as $4s,4p$ shells have to be included to the calculation.  Their
relative importance can be catched only in the calculation which must account
for electronic correlation.  In turn, $3s$ and $3p$ orbitals (and those with
``lower'' $n$) have essentially different space localization
(4.5 times smaller averaged radii) and too large energy denominator $\sim 60$ a.u.  
Therefore, we do not consider these shells in the paper, particularly, because their explicit
treatment dramatically increase the computational effort.

\begin{table}[!h]
\caption{
averaged radii, $\langle r \rangle$, last maximum position, r$_{max}$, and
orbital energies, $\varepsilon_{\rm orb}$, of Eu$^{++}$ spin-orbit averaged spinors.
}
\label{TOrbitalParameters}
\begin{tabular}{ c  c  c  c  c  c  c }
\hline\hline
 orbital ~~ &
 $\langle r \rangle$, a.u. ~~&
 r$_{max}$, a.u. ~~&
 $\varepsilon_{\rm orb}$, a.u. ~~\\\hline
 $3s$ &
 0.2 &
 0.3 &
{}-68.0\\
 $3p$ &
 0.2 &
 0.3 &
{}-57.7\\
 $3d$ &
 0.2 &
 0.3 &
 {}-43.6\\\hline
 $4s$ &
 0.6 &
 0.5 &
 {}-15.1\\
 $4p$ &
 0.6 &
 0.5 &
 {}-11.5\\
 $4d$ &
 0.7 &
 0.6 &
 {}-6.5\\
 $4f$ &
 0.9 &
 0.6 &
 {}-1.0\\
 $5s$ &
 1.4 &
 1.3 &
 {}-2.6\\
 $5p$ &
 1.6 &
 1.4 &
 {}-1.6\\\hline\hline
\end{tabular}
\end{table}

In order to estimate contributions from different shells we have performed a
series of calculations at the CCSDT level with the ROHF reference calculated at
zero external electric field.  In these calculations we have frozen different
orbitals, i.e., have forbidden their spin and space polarization. For every
calculation we have also decomposed $K$ on contributions going from the $s-p$
($K_{s-p}$), $p-d$ ($K_{p-d}$), and $f-d$ ($K_{f-d}$) mixings.  In paper
\cite{Buhmann:02}, the Gd$^{+++}$ system electronically equivalent to Eu$^{++}$
was studied where the scheme of the $s-p$ mixing with excitations to unoccupied
$d$ states was considered.  To analyze importance of other possible correlation
schemes of the $s-p$ mixing in Eu$^{++}$ we have performed all these calculations in
two basis sets:
(i) our standard basis set Lbas (which has been used for previous calculations)
and 
(ii) the Lbas\_nvd basis set which was derived from Lbas by keeping all the s,
p, f and only one contracted 4d function (taken from ROHF) and excluding all the
other $d$ functions.  This means that no virtual $d-$orbitals will appear in
Hartree-Fock calculations with the Lbas\_nvd basis (addition ``nvd'' means ``No
Virtual $d$'').  In such a way we have excluded virtual $d-$orbitals from the
calculations. The results of the calculations are given in table
\ref{TMechanisms}.

\begin{table*}[!h]
\caption{
Calculated $K$ values and its components  using the ROHF-CCSDT method 
(The ROHF-reference is taken from zero \Eext\ field calculation to prevent space
polarization of the orbitals to be frozen).
}
\label{TMechanisms}
\begin{tabular}{c  c  c  c  c  c  c | c  c  c  c}
\hline\hline
\centering  \# &
\centering  Active orbitals &
\centering Frozen orbitals &
\multicolumn{4}{c}{\centering  Lbas} &
\multicolumn{4}{c}{\centering  Lbas\_nvd}\\\hhline{~~~--------}
~
 &
~
 &
~
 &
 $K$ &
 $K_{s-p}$ &
 $K_{p-d}$ &
 $K_{f-d}$ &
 $K$ &
 $K_{s-p}$ &
 $K_{p-d}$ &
 $K_{f-d}$\\\hline
 1 &
 all &
 {}- &
 {}-4.7 &
 {}-3.4 &
 0.3 &
 {}-1.7 &
 1.6 &
 1.8 &
 0.0 &
  {}-0.3\\
 2 &
 $4f-$ only &
 $4s4p4d5s5p$ &
 {}-1.8 &
 0.0 &
 0.0 &
 {}-1.8 &
 0.0 &
 0.0 &
 0.0 &
  0.0\\
 3 &
 $5s, 4f$ &
 $4s4p4d\_5p$ &
 {}-0.6 &
 1.2 &
 0.0 &
 {}-1.8 &
 0.8 &
 0.8 &
 0.0 &
  0.0\\
 4 &
 $5p, 4f$ &
 $4s4p4d5s\_$ &
 {}-1.3 &
 {}-0.9 &
 0.8 &
 {}-1.2 &
 2.1 &
 2.1 &
 0.0 &
  0.0\\
 5 &
 $5s\&5p, 4f$ &
 $4s4p4d\_\_$ &
 {}-0.3 &
 0.2 &
 0.8 &
 {}-1.3 &
 2.0 &
 2.0 &
 0.0 &
  0.0\\
 6 &
 $4d\&5s\&5p, 4f$ &
 $4s4p\_\_\_$ &
 {}-1.9 &
 {}-0.8 &
 0.6 &
 {}-1.7 &
 1.3 &
 1.5 &
 0.0 &
  {}-0.3\\
 7 &
 $4s, 4f$ &
 $\_4p4d5s5p$ &
 {}-3.7 &
 {}-1.9 &
 0.0 &
 {}-1.8 &
 0.0 &
 0.0 &
 0.0 &
  0.0\\
 8 &
 $4p, 4f$ &
 $4s\_4d5s5p$ &
 {}-3.2 &
 {}-1.1 &
 {}-0.3 &
 {}-1.8 &
 0.1 &
 0.1 &
 0.0 &
  0.0\\
 9 &
 $4s\&4p, 4f$ &
 $\_\_4d5s5p$ &
 {}-5.4 &
 {}-3.3 &
 {}-0.3 &
 {}-1.8 &
 0.1 &
 0.1 &
 0.0 &
  0.0\\
 10 &
 4s\&4p\&4d, 4f &
 \_\_\_5s5p &
 {}-6.8 &
 {}-3.8 &
 {}-0.7 &
 {}-2.2 &
 {}-0.1 &
 0.1 &
 0.0 &
  {}-0.2\\
 11 &
 $4s\&4p\&5s\&5p, 4f$ &
 $\_\_4d\_\_$ &
 {}-2.5 &
 {}-1.9 &
 0.6 &
 {}-1.3 &
 2.3 &
 2.3 &
 0.0 &
  0.0\\
 12 &
 $4d, 4f$ &
 $4s4p\_5s5p$ &
 {}-2.7 &
 0.0 &
 {}-0.4 &
 {}-2.2 &
 {}-0.2 &
 0.0 &
 0.0 &
  {}-0.2\\\hline\hline
\end{tabular}
\end{table*}

One can make many fruitful conclusions from this table, but we shall point out only
at several of them:\\
{\bf (i)} $K_{s-p}$ contribution from the spin and space polarization (below we
call the simultanious effect as just the polarization) of $5s$ and $5p$
orbitals.  To extract a ``clean'' contribution from polarization of the $5s$
orbital we have frozen $4s,4p,4d$ and $5p$ orbitals, thus, only the $5s$ and
$4f$ occupied shells were included in the correlation calculation (see line 3 in
the table).  The $K_{s-p}$ value for this case in Lbas is $+1.2$.  The case with
excluded virtual d-basis functions (using Lbas\_nvd basis) gives K$_{s-p}=+0.8$.
Therefore, one can conclude that the mechanism of polarization of the 5s-orbital
including intermediate excitation into virtual d-orbitals (being large in lowest
PT orders as is shown in \cite{Buhmann:02}) is supressed in the higher PT orders
and, therefore, can not be considered as the leading one.  Moreover, as the 4d
shell is frozen in this calculation the contribution originates from the
spin-polarization (by seven electrons occupying 4$f_{\uparrow}$-orbitals) of the
occupied 5s-orbital directly and indirectly, by means of the virtual p states,
that is complemented by the space polarization of these 5s states into virtual
p-states.

``Clean'' contribution from polarization of the 5p-orbitals (below will write
just as contributions of 5p for brevity, etc.) to $K _{s-p}$ is $-0.9$ (that
includes intermediate virtual s, p, and d functions, see line 4).  Note that the
nvd mechnism gives $K _{s-p}=+2.1$. Both polarizations of
5s and 5p orbitals give significant contributions, however, these terms have
opposite signs and simultaneous correlation of 5s and 5p orbitals (see line 5 of
the table) results in almost negligible $K_{s-p}$ value, $+0.2$ (note that the
sum of the partial 5s ($+1.2$) and 5p ($-0.9$) contributions is $+0.3$,
therefore, they are practically additive).  It should be noted that the
nvd$-$mechanism gives 
$K_{s-p}=+2.0$, therefore, it looks like that there is almost exact compensation
between contributions with intermediate virtual $s, p$ and virtual $d$ states.
Additional inclusion of $4d$ orbitals in the correlation calculation (line 6)
leads to decrease of $K_{s-p}$ by 1.0.

{\bf (ii)} $K_{s-p}$ contribution from the polarization of $4s$ and $4p$
orbitals.  From lines 7, 8 and 9 one can see that the individual polarization
contributions to $K _{s-p}$ from the $4s$ and $4p$ orbitals are slightly higher
(by absolute value) than the corresponding $5s$ and $5p$ contributions, and,
in turn, they have the same signs which leads to big final $K _{s-p}$; also both
$4s$ and $4p$ polarizations are mainly due to the ``virtual $d$'' mechanism and
they are also almost additive.  Additional inclusion of 4d-orbitals in the
correlation calculation (line 10) leads to decrease of  $K_{s-p}$ by 0.5.
At last, it is the polarization of $4s$ and $4p$ orbitals that gives the {\em
leading} contribution to the final $K _{s-p}$ and total $K$ values.

{\bf (iii)} Simultaneous correlation of $4s$, $4p$, $5s$, $5p$ (and, of course,
$4f$) orbitals (line~11) leads to certain decrease of $K _{s-p}$ by absolute
value with respect to the sum of $4s, 4p$ and $5s, 5p$ contributions.  Here we
have new types of ``interfering polarization contributions'' e.g., between $4s$
and $5p$, $4p$ and $5s$, etc.  Additional inclusion of 4d-orbitals in the
correlation calculation (line~1) leads to further decrease of the $K_{s-p}$
value by 1.5.

{\bf (iv)} $K_{f-d}$ notably depends on the $5p$ orbital (compare lines~2 and
4).  One can expect that this is mainly due to the space $5p-5d$ polarization
(together with the spin-polarization to intermediate virtual d states) that
decrease the contribution from the direct (lowest order) space $4f-5d$
polarization.

{\bf (v)} The polarization of $5p$ to virtual $d$ (lines 2, 4) and of $4d$ to
virtual $p$ orbitals (lines 2, 12), as well as the small term with polarization
of $4p$ to virtual $d$ (lines 2, 8) have the opposite signs (and small
magnitudes) resulting in a negligible final value of $K_{p-d}$ (line 1).

 One important note should be made for the $K_{f-d}$ contribution.  The
 amplitude of the $f$ function is very small at the core region of Eu$^{++}$.
 Therefore, it is  important to take into account the electric field screening
 effect from the core shells. We have calculated that neglecting these screening
 effects leads to 25\% overestimation of the $K_{f-d}$ contribution.
 Also, is should be noted that all the core shells give screening effects, i.e.\
 not only $1s$, but $2s-3d$ give essential contribution there as well. 

It is clear from table \ref{TMechanisms} that mechanisms of forming the final
enhancement factor value $K$ are very complicated, and one should consider many
orders of perturbation theory by interelectronic Coulomb interaction (as
expected from the qualitative discussion in section ``Methods''). As is stressed
above, it is the coupled clusters theory which includes many orders of
perturbation theory that we have chosen for our calculation.
This is the only method that allowed us to attain convergent results for
such a complicated problem as evaluating $K$ in lanthanides.
%

\section{Density functional and M\o ller-Plesset estimates}

As was mentioned above this paper is the first one in our studies of \Eeff\ on
Eu$^{++}$ having in mind to describe the effective state of Eu in the crystal
Eu$_{0.5}$Ba$_{0.5}$TiO$_{3}$ as our final goal. Since in solid-state
calculations one cannot use such methods as coupled clusters we have calculated
$K$ using the M\o ller-Plesset perturbation theory (see table \ref{TCorrMethod}
above) and different popular exchange-correlation functionals.
Although the second order M\o ller$-$Plesset perturbation theory with the
UHF-reference is in a good agreement with our final value of $K$ (and even
individual $s-p$, $p-d$ and $f-d$ contributions are so as well)
for the system under 
consideration, however we cannot consider the UHF$-$MP2 values as reliable enough
because the ROHF$-$based MP2, as well as the UHF$-$ and ROHF$-$based MP3 and MP4
values are seriously divergent from them demonstrating instability of the MP
series. Thus, one can try to use MP2 (mainly for many-atomic systems) but with
great caution.

Unfortunately, there are no reliable theoretical criteria to choose the
most appropriate DFT exchange-correlation functional versions
for a problem of the considered type because it is impossible to perform a series 
of successive DFT calculations with
consistent increase of the level of accuracy of theory to achieve convergence as
it can be done, at least formally, in the framework of the explicitly-correlated
ab initio methods (see above).  Therefore the only a way to choose density
functional is to ``calibrate'' it comparing to a high-level correlation
calculations. In table \ref{TDFTResults} we present the calculated $K$ values
using different exchange-correlation functionals. One should note that
$K_{s-p}$, $K_{p-d}$ and $K_{f-d}$ have the same weights as in the case of
CCSDT.

\begin{table}[!h]
\caption{
Calculated $K$ values using popular exchange-correlation functionals.
}
\label{TDFTResults}
\begin{tabular}{ l  c }
\hline\hline
Functional &
    $K$ \\
\hline
  PBE \cite{Perdew:96} &
    {}-3.7\\
  TPSS \cite{Tao:03} &
    {}-3.8\\
  B3LYP \cite{b3lyp} &
    {}-2.9\\
  PBE0 \cite{pbe0} &
    {}-2.7\\
\hline\hline
\end{tabular}
\end{table}

\section{Conclusion}

The Eu$^{++}$ cation in an external electric field has been considered as the
simplest important model that simulates effective state of europium in our
studies of the EBTO crystal properties.
The calculated enhancement factor is $K=-4.6$. 
It is shown that this value is not well determined by even the lowest four orders of 
the many-body pertubation theory by the Coulomb operator, so the coupled-cluster 
expansion for the wave function is important to attain a convergence for this value.
The other exploited methods including the RASSCF and CI ones did not allow us 
to attain the convergence on $K$ in a reasonable time using available computer resources. 
The main contribution to $K$ originates from the spin-space
polarization of $s$ and $p$ occupied orbitals.

\section{Acknowledgments}
This work was supported by RFBR grant 09--03--01034 and, partially,
10--03--00727.  L.S.\ is grateful 
to the Dmitry Zimin ``Dynasty'' Foundation and for grant from Russian Science Support
Foundation.  AT is grateful to UNSW for support within the Professorial visiting
fellowship program.


\end{document}